\documentclass[11pt]{article}

\usepackage[final]{acl}

\usepackage{times}
\usepackage{latexsym}
\usepackage{hyperref}
\usepackage[T1]{fontenc}

\usepackage[utf8]{inputenc}

\usepackage{microtype}

\usepackage{inconsolata}

\usepackage{graphicx}

\usepackage{adjustbox}
\usepackage{array}
\usepackage{xspace}
\usepackage{amsmath}
\usepackage{booktabs}
\usepackage{multirow}
\usepackage{xcolor}

\newcolumntype{R}[2]{%
    >{\adjustbox{angle=#1,lap=\width-(#2)}\bgroup}%
    l%
    <{\egroup}%
}
\newcommand{\name}{{\tt ColBERT-Att}\xspace}
\newcommand{\comment}[1]{}

\DeclareMathOperator*{\argmax}{argmax}
\definecolor{platinum}{rgb}{0.9, 0.89, 0.89}

\title{{\name}: Late-Interaction Meets Attention for Enhanced Retrieval}

\author{Raj Nath Patel \\
  Huawei Research Center \\
  Dublin, Ireland \\
  \texttt{raj.nath.patel@huawei.com} \\ \And
  Sourav Dutta \\
  Huawei Research Center \\
  Dublin, Ireland \\
  \texttt{sourav.dutta2@huawei.com} \\}

\begin{document}
\maketitle

\begin{abstract}
    Vector embeddings from pre-trained language models form a core component in Neural Information Retrieval systems across a multitude of knowledge extraction tasks. The paradigm of {\em late interaction}, introduced in ColBERT, demonstrates high accuracy along with runtime efficiency. However, the current formulation fails to take into account the attention weights of query and document terms, which intuitively capture the ``importance'' of similarities between them, that might lead to a better understanding of relevance between the queries and documents. This work proposes \name, to explicitly integrate {\em attention mechanism} into the late interaction framework for enhanced retrieval performance. Empirical evaluation of \name depicts improvements in recall accuracy on MS-MARCO as well as on a wide range of BEIR and LoTTE benchmark datasets.
\end{abstract}

\section{Introduction \& Background}
\label{sec:intro}

{\em Semantic similarity} or relevance between queries and documents using high-dimensional
dense vector representations of texts, from large language models, has become ubiquitous in Information Retrieval (IR) and also forms a core component in Retrieval Augmented Generation
(RAG)~\cite{lewis2020retrieval}.
Retrieval of relevant documents or information has transitioned from lexical and text matching (e.g., BM25~\cite{robertson1995okapi}) to semantic retrieval based on neural models (e.g., ColBERT~\cite{khattab2020colbert}) capturing the semantics and context of information contents.

Traditional systems like BM25 using simplistic text matching (e.g., TF-IDF~\cite{sparck1972statistical, salton1988term} measure) based on {\em sparse encoding} and corpora statistics, failed to capture similarities beyond surface form equivalence.
SPLADE~\cite{formal2021splade} provides an explicit sparsity regularization and a log-saturation effect for obtaining term weights, eliminating the need of statistics and hyper-parameters. 
However, it suffers from language dependency along with expensive memory and compute requirements.
BM42~\cite{vasnetsov2024bm42} aims to combine the strengths of lexical matching and attention mechanism from language models. The use of document token attention weights as a proxy to documents term importance was shown to be effective.

Orthogonally, neural IR methods~\cite{karpukhin2020dense,qu2021rocketqa,zhan2020repbert,gao2021coil,luan2021sparse} allow semantic and contextual matching by encoding queries and documents into {\em single-vector dense representations} from language models to gauge the relevance between queries and documents. {\em Late interaction} strategy used in ColBERT, produces {\em multi-vector representations}
at token or subtoken level granularity, and computes fine-grained relevance scores between query and document tokens using efficient and scalable token-level computations.

Specifically, the original ColBERT architecture computes the maximum cosine similarity (i.e., {\em MaxSim} operation) between a query token and all the document tokens, using the token-level dense representations. Finally, a summation of the maximum similarities over all the query tokens is computed to obtain the final relevance scores between query and documents. 
Consider, $\mathcal{E}_{q_i}$ and $\mathcal{E}_{d_j}$ to denote the embeddings of the $i^{th}$ token of query $\mathcal{Q}$ and the $j^{th}$ token in document $\mathcal{D}$ respectively. Mathematically, the score is thus computed as, 
\begin{align}
\label{eq:colbert}
    \mathcal{S}_{\mathcal{Q},~\mathcal{D}} = \sum_{q_i \in \mathcal{Q}} \max_{d_j \in \mathcal{D}} \mathcal{E}_{q_i} \cdot \mathcal{E}_{d_j}
\end{align}
%
ColBERT has been shown to perform extremely well on retrieval tasks by leveraging the semantic expressiveness of language model embeddings, along with the ability to pre-compute document representations for speeding up query processing latency. To improve upon the accuracy and memory footprint of the late interaction architecture, vector compression techniques and denoising training strategy were proposed in ColBERTv2~\cite{santhanam2022colbertv2}. The use of centroid interaction and pruning approach were introduced in ColBERTv2$_{\text{\small{PLAID}}}$~\cite{santhanam2022plaid}, to improve the search latency with comparable performance. Rotary positional embeddings~\cite{su2024roformer} and advanced activation functions for enhanced retrieval performance were recently incorporated in ModernColBERT~\cite{GTE-ModernColBERT}.

\noindent {\bf Motivation.} Observe that the {\em MaxSim} operation in ColBERT does not explicitly factor in the importance of either the query or document terms. As such, all term matches between query and documents are considered to be of equal importance, which is not necessarily true and might degrade the overall performance. Although, vector embeddings obtained from language models implicitly leverage attention weights, we posit that explicit incorporation of terms importance via attention mechanism within the relevance score computation could be beneficial. As an intuition of how attention weights can enable better understanding of {\em query-document relevance}, consider the following toy example. 

\hspace*{-4mm}\fcolorbox{platinum}{platinum}{
\parbox{0.475\textwidth}{
Assume a query and candidate documents as: 

{\bf $\mathcal{~Q}$}: {\tt Who is going to \underline{study}?} \\
{\bf $\mathcal{D}_1$}: {\tt Alice is walking to \underline{school}.} \\ 
{\bf $\mathcal{D}_2$}: {\tt Bob is going to \underline{buy apples}.} \\
{\bf $\mathcal{D}_3$}: {\tt Only studying makes Jack a \underline{dull boy}.} \\
Here, we highlight probable key terms (underlined) within the query and documents, which would ideally have higher attention weights from a contextual language model.

\hspace*{3mm} Observe, the phrase {\em ``is going to''} in $\mathcal{Q}$ has a high (embedding based cosine) similarity with $\mathcal{D}_2$, since it is present in both the texts. However the attention weights of these terms are relatively low in $\mathcal{Q}$, as they are not important to the overall context and intent of the query. Thus, incorporation of query term attention would effectively and accurately reduce the relevance between $\mathcal{Q}$ and $\mathcal{D}_2$. 

\hspace*{3mm} Similarly, the term {\em ``studying''} having low attention weight in document $\mathcal{D}_3$ would diminish its relevance to the query, in spite of a high similarity match with the term ``study'' in $\mathcal{Q}$.

\hspace*{3mm} On the other hand, although the terms {\em ``study''} and {\em ``school''} in $\mathcal{Q}$ and $\mathcal{D}_1$ resp. are related (having moderate cosine similarity), the high query and document term attention weights would {\em boost} the relevance score between the query and document -- thus accurately retrieving $\mathcal{D}_1$ for query $\mathcal{Q}$.
}
} 

Thus, incorporation of query and document attention weights within the late-interaction framework could potentially improve the overall retrieval performance, as proposed here in \name .

\section{\name Training}
\label{sec:method}

Consider a query $\mathcal{Q}$ and a document $\mathcal{D}$ to be composed of $n$ and $m$ tokens respectively. Thus, $\mathcal{Q} = \{q_1, q_2, \cdots, q_n\}$ and $\mathcal{D} = \{d_1, d_2, \cdots, d_m\}$. The relevance score of the document to the query ($\mathcal{S}_{\mathcal{Q},~\mathcal{D}}$) is then computed as,
\begin{align}
\label{eq:colbert-att}
    \mathcal{S}_{\mathcal{Q},~\mathcal{D}} = \sum_{i=1}^{n} e^{~\mathcal{A}_{q_i}} \cdot \max_{j=1}^m (\mathcal{E}_{q_{i}} \odot \mathcal{E}_{d_{j}}) \cdot (e^{~\mathcal{A}_{d_{w}}})^{^\delta}
\end{align}
where, $\mathcal{E}$ and $\mathcal{A}$ represents the corresponding vector embeddings and attention weights respectively.
Further, $\odot$ denotes the cosine similarity operator, and the document token that depicts the highest similarity to the query token $q_i$ is represented as $d_w = \argmax_{j \in [1, \cdots, m]} ~(\mathcal{E}_{q_{i}} \odot \mathcal{E}_{d_{j}})$. 
Finally, $\delta$ signifies a document length based {\em attention weight regularizer}, which we discuss later in this section.

In other words, we augment the {\em MaxSim} operation of Eq.~\eqref{eq:colbert} with the corresponding query token attention weight along with the associated document token attention weight that depicts the highest (cosine) similarity to the query token. Since the attention weights are typically small values, we accentuate their values (and relative differences) by taking the exponent. Following the original framework of ColBERT, we consider the queries to comprise $32$ tokens and documents to be represented by $300$ tokens (including special and mask tokens) for most datasets unless specified otherwise.

To obtain our \name model, we trained ColBERTv2$_{\text{\small{PLAID}}}$ (obtained from \url{https://github.com/stanford-futuredata/ColBERT}) with the modified objective of Eq.~\eqref{eq:colbert-att} (with $\delta=1$) using the official train queries and corresponding positive/negative triples of MS-MARCO dataset. Training was performed for 1M steps with default parameter settings, and the best checkpoint was considered. The document token embeddings and attention weights are computed offline and stored as a pre-processing step, while the query token embeddings and attentions are obtained on-the-fly during inference. Observe, that the attention weights are technically free (in terms of compute), as they are an inherent artifact of the encoding process -- thus has {\em no impact on inference latency}. 

\begin{table}[t]
\centering
    \resizebox{0.95\columnwidth}{!}{
    \begin{tabular}{lccc}
    \toprule
    ~ & {\bf R@50} & {\bf R@100} & {\bf R@1K} \\
    \midrule
    \text{ColBERTv2$_{\text{\tiny{PLAID}}}$} & 86.76 & 91.36 & 97.58 \\
    \textit{\name} & \bf{86.78} & \bf{91.54} & \bf{97.64} \\
    \bottomrule
    \end{tabular}
    }
    \caption{\small Results on {\bf MS-MARCO} Passage Ranking dev set.}
    \label{tab:msmarco}
\end{table}

\begin{table*}[t]
  \centering
  \vspace*{-1mm}
  \resizebox{0.85\textwidth}{!}{
    \begin{tabular}{lcccccc}
    \toprule
    ~ & \multicolumn{6}{c}{{\bf LoTTE Search Test Queries} ({\em Success@5})} \\
    \hline
    \hline
    ~ & {\textbf{ColBERT}} & {\textbf{BM25}} & {\textbf{ANCE}} & {\textbf{RocketQAv2}} & {\textbf{ColBERTv2$_{\text{\tiny{PLAID}}}$}} & {\textbf{\name}} \\
    \midrule
    \textit{Lifestyle} & 80.2 & 63.8 & 82.3 &  82.1 &  84.3   &  \bf{84.9}   \\
    \textit{Science}   & 53.6 &  32.7 & 53.6 & 55.3 &  56.6   &  \bf{56.9}  \\
    \textit{Writing}   & 74.7 & 60.3 & 74.4 & 78.0 & 79.5 & \bf{80.2}   \\
    \textit{Recreation}  & 68.5 &  56.5 & 64.7 & 72.1 & 71.6 &  \bf{72.3}   \\
    \textit{Technology}   & 61.9 & 41.8 & 59.6 & 63.4 & 66.1  &  \bf{67.8}   \\\hline
    \textit{Weighted Av.}   & 68.82 & 52.73 & 72.88 & 71.42 & 72.7 & \bf{73.5}  \\
    \toprule
    ~ & \multicolumn{6}{c}{{\bf LoTTE Forum Test Queries} ({\em Success@5})} \\
    \hline
    \hline
    ~ & {\textbf{ColBERT}} & {\textbf{BM25}} & {\textbf{ANCE}} & {\textbf{RocketQAv2}} & {\textbf{ColBERTv2$_{\text{\tiny{PLAID}}}$}} & {\textbf{\name}} \\
    \midrule
    \textit{Lifestyle}   & 73.0 &  60.6 & 73.1 & 73.7 &  76.7   &  \bf{77.2}   \\
    \textit{Science}   & 41.8 & 37.1 & 36.5 & 38.0 & 46.1   &  \bf{46.5}   \\
    \textit{Writing}   & 71.0 & 64.0 &  68.8 & 71.5 & 75.7   &  \bf{77.1}   \\
    \textit{Recreation}   & 65.6 &  55.4 & 63.8 & 65.7 &  70.6   &  \bf{70.7}   \\
    \textit{Technology}   & 48.5 & 39.4 &  46.8 & 47.3 & 53.2   &  \bf{54.3}   \\\hline    \textit{Weighted Av.} & 59.94 & 51.27 & 57.76 & 59.20 & 64.4  &  \bf{65.1} \\
    \bottomrule
    \end{tabular}
    }
  \caption{Evaluation results on {\bf LoTTE} {\em Search and Forum} datasets. {\small (Best results are presented in {\bf bold}.)}}
  \label{tab:lotte}
\end{table*}

\begin{table*}[t]
  \centering
  \resizebox{0.9\textwidth}{!}{
  \begin{tabular}{lcccccccc}
  \toprule
  ~ & \multicolumn{8}{c}{{\bf BEIR Search Tasks} ({\em nDCG@10})} \\
  \hline
  \hline
    ~ & {\textbf{ColBERT}}  & {\textbf{DPR-M}} & {\textbf{ANCE}} & {\textbf{MoDIR}} & {\textbf{TAS-B}} & {\textbf{RocketQAv2}} & {\textbf{ColBERTv2$_{\text{\tiny{PLAID}}}$}}  & {\textbf{\name}} \\
    \midrule
    \textit{FiQA}  & 31.7 &  27.5 & 29.5 &  29.6 &  30.0 & 30.2 &  \bf{35.1}   &  \underline{34.8}  \\
    \textit{NFCorpus} & 30.5 & 20.8 & 23.7 & 24.4 & 31.9 & 29.3 & \underline{33.0}   &  {\bf 33.1} \\
    \textit{NQ} & \bf{52.4} & 39.8 &  44.6 & 44.2 &  46.3 & \underline{50.5} & 48.8  &  49.0 \\
    \textit{HotpotQA} & 59.3 &  37.1 & 45.6 &  46.2 & 58.4 & 53.3 & \bf{66.1}  & \underline{65.9} \\
    \toprule
    ~ & \multicolumn{8}{c}{{\bf BEIR Semantic Relatedness Tasks} ({\em nDCG@10})} \\
    \hline
    \hline
    ~ & {\textbf{ColBERT}}  & {\textbf{DPR-M}} & {\textbf{ANCE}} & {\textbf{MoDIR}} & {\textbf{TAS-B}} & {\textbf{RocketQAv2}} & {\textbf{ColBERTv2$_{\text{\tiny{PLAID}}}$}}  & {\textbf{\name}} \\
    \midrule
    \textit{ArguAna} & 23.3 & 41.4 &  41.5 & 41.8 & 42.7 &  {\bf 45.1} & 42.06  & \underline{44.3}  \\
    \textit{SciFact} & {\bf 67.1} & 47.8 & 50.7 &  50.2 &  64.3 & 56.8 & \underline{67.1}  & 66.2  \\
    \textit{SCIDOCS} & \underline{14.5} & 10.8 & 12.2 & 12.4 & {\bf 14.9} & 13.1 & 14.3  & \underline{14.5}  \\
    \textit{Quora} & \underline{85.4} & 84.2 & 85.2 & \bf{85.6} & 83.5 & 74.9 & 84.9 & \underline{85.4} \\
    \textit{FEVER} &  \underline{77.1} & 58.9 & 66.9 & 68.0 & 70.0 & 67.6 & 76.53 &  \bf{77.4}  \\
    \textit{C-FEVER} & 18.4 &  17.6 & 19.8 & \underline{20.6} & {\bf 22.8} & 18.0 & 16.7 &  17.6   \\
    \bottomrule
    \end{tabular}
    }
  \caption{Evaluation results on {\bf BEIR} {\em Search and Semantic Relatedness} datasets. {\small (Best results are presented in {\bf bold}, while second-best results are \underline{underlined}. For {\em ArguAna}, the query was represented with $300$ tokens, as used in the literature.)}}
  \label{tab:beir}
\end{table*}

\begin{table*}[ht!]
  \centering
  \resizebox{\textwidth}{!}{
  \begin{tabular}{lccccc}
    \multicolumn{5}{c}{~} & \hspace*{5mm}\multirow{7}{*}{\fbox{\includegraphics[width=0.9\columnwidth]{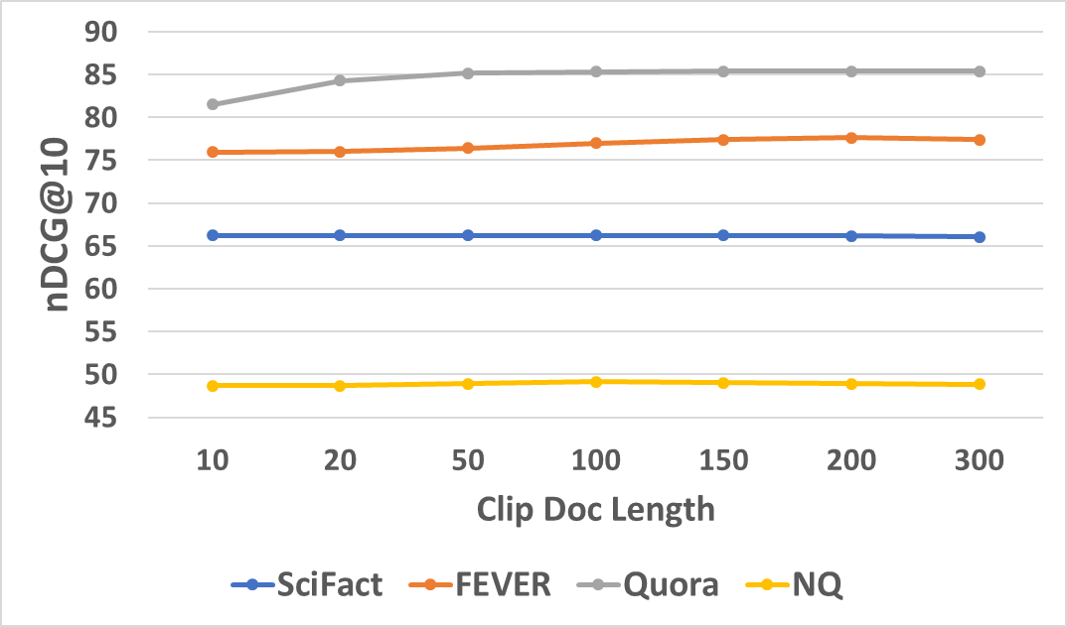}}} \\
    \cmidrule{1-5}
    ~ & \textbf{No Attn.} & \textbf{Only $\mathcal{A}_q$} & \textbf{Only $\mathcal{A}_D$} & \textbf{\name} & \\
    \cmidrule{1-5}
    \textit{Lifestyle} & 84.72 & 84.42 & 84.87 & \bf{84.87} &   \\
    \textit{Science}  & 56.89 & 57.05 & \bf{58.02} &  56.89 &   \\
    \textit{Writing}  & 79.08 & 79.65 & 79.74 & \bf{80.21} &  \\
    \textit{Recreation}  & 71.86 & 72.29 & 72.19 & \bf{72.29} & \\
    \textit{Technology}  &  66.78 & 66.94 & 67.62 & \bf{67.79} & \\
    \cmidrule{1-5}
    \multicolumn{6}{c}{~} \\
    \multicolumn{5}{c}{\bf (a)} & \hspace*{5mm}{\bf (b)}
    \end{tabular}
    }
  \caption{Ablation study for \name: {\bf (a)} Results of Success@5 with different {\em attention inclusion} on {\bf LoTTE}, and {\bf (b)} Impact of {\em attention regularizer} ($\delta$) in $\mathcal{A}_D$ with varying document length clipping on nDCG@10 for {\bf BEIR}.}
  \label{tab:attn}
\end{table*}

{\bf Attention Weight Regularizer.} It is important to note that document length plays a significant effect on the values of the attention weights, with longer documents demonstrating lower token attention weights compared to shorter ones. Thus, substantial difference between the attention weights encountered during training of \name and those during inference would introduce discrepancies and might degrade the overall retrieval performance. In fact, the average document length for MS-MARCO (used for training) is around $55$, while the average document lengths range from $10$ (for Quora) to $230$ (in NFCorpus) across other datasets in the BEIR evaluation benchmark.

To alleviate the above, in the formulation of Eq.~\eqref{eq:colbert-att} we introduce the {\em attention weight regularizer} ($\delta$), defined as $\delta = \min(1, doc\_len / l)$. We empirically set the document length clipping hyper-parameter $l=150$, discussed later in Section~\ref{sec:expts}. Effectively, this regularizer aims to scale-down high attention weights (for shorter documents), while keeping the original values for others (using $\min$).

For model training we used $2$ NVIDIA A100 GPUs (with $80$ GB each), while inference was conducted on an NVIDIA Quadro RTX GPU ($16$ GB).

\comment{
\verb|Important NOTE for Sourav:| the results are with PLAID settings as mentioned in the \href{https://github.com/stanford-futuredata/ColBERT/issues/410}{comment} . It compares with the \href{https://arxiv.org/pdf/2205.09707}{PLAID paper's} reported results. 
}

\section{Empirical Results}
\label{sec:expts}

We evaluate our proposed approach on a wide variety of retrieval tasks from open-source benchmark datasets. Specifically, we compare the performance across several existing methodologies using the {\bf MS-MARCO}~\cite{nguyen2016ms} (dev split), {\bf BEIR}~\cite{thakur2021beir} (search and semantic relatedness tasks), and {\bf LoTTE}~\cite{santhanam2022colbertv2} (search and forum) datasets. In terms of evaluation measures, we report {\em Recall@k}, {\em nDCG@10}, and {\em Success@5} respectively for the different benchmarks, as shown in literature.

From Table~\ref{tab:msmarco}, we observe that \name achieves a performance improvement of $0.2$\% on Recall@100 even for the challenging {\bf MS-MARCO} dataset, wherein baseline methods perform quite high. This constitutes {\em in-domain} evaluation, as the model has been trained on MS-MARCO, and here we set $\delta=1$ during inference.

To showcase the efficacy of our framework on {\em out-of-domain} datasets, we evaluate on {\bf LoTTE}, that focuses on natural search queries on documents with long-tailed topics, unlike open-ended QA of the BEIR dataset. From Table~\ref{tab:lotte}, we observe \name to consistently outperform the existing approaches on all the datasets with an avg. improvement of $\sim 1$\% on the Success@5 metric.

For completeness, we also evaluate the different methods on a range of {\bf BEIR} datasets spanning search and semantic retrieval tasks as presented in Table~\ref{tab:beir}. The baseline results reported are from~\cite{santhanam2022colbertv2}, while ColBERTv2$_{\text{\small{PLAID}}}$ results are obtained by executing the code repository available at \url{github.com/stanford-futuredata/ColBERT}. Here, we observe \name to perform better (on most datasets) than the original  ColBERTv2$_{\text{\small{PLAID}}}$ 
model (which uses the MaxSim without any attention weights). In fact, on {\em ArguAna}, we obtain a significant gain of around $2$\%. Overall, \name is seen here to be comparable to the other baselines. 

Observe that ColBERTv2 has been shown to perform better than all existing baselines (including SPLADE)~\cite{santhanam2022colbertv2}. Due to the unavailability of the original code of ColBERTv2, our current implementation is based on ColBERTv2$_{\text{\small{PLAID}}}$ (which performs slightly worse compared to ColBERTv2)~\cite{santhanam2022plaid}. 
Overall, we present that within the current framework, inclusion of attention weights tends to improve the accuracy for different retrieval tasks on multiple benchmark datasets. Incorporation of our objective in {\em ModernColBERT} framework, an interesting direction of future study, provides hope of achieving state-of-the-art retrieval results.

{\bf Ablation Study.} Table~\ref{tab:attn}(a) depicts how the inclusion of both the query and document attention weights in the {\em MaxSim} operation of Eq.~\eqref{eq:colbert-att} provides the best performance for \name.

To evaluate the effect of {\em attention regularizer} ($\delta$) in \name, we vary the document length clipping hyper-parameter $l$ and report the observed results in Table~\ref{tab:attn}(b). We observe that this strategy can efficiently handle document length (i.e., attention weight value) mismatches between training and inference -- leading to an impressive $5$\% nDCG@10 improvements on Quora (having $5\times$ lower avg. document length compared to MS-MARCO training data). Overall, \name is seen to be quite robust across a wide range of values, and we set $l=150$ for our experimental setup.

\section{Conclusion}
\label{sec:conc}

This work presented a novel framework, \name, that explicitly integrates the {\em late interaction} mechanism with attention weights. We show that this incorporation of query and document term importance through attention weights within the {\em MaxSim} operation along with document length based attention regularizer, provides improved accuracy on diverse retrieval tasks from multiple benchmark datasets. 




\bibliography{custom}

\end{document}